\documentclass[preprint,showpacs,amsmath,amssymb]{revtex4}
\usepackage{graphicx}
\usepackage{dcolumn}% Align table columns on decimal point
\usepackage{bm}% bold math

\begin{document}

\title{The role of $N^{*}(1535)$ in $\eta '$ production}

\author{Xu Cao$^{1,3}${\footnote{Email: caoxu@impcas.ac.cn}}, Xi-Guo Lee$^{1,2}${\footnote{Email: xgl@impcas.ac.cn}}}

\affiliation{1. Institute of Modern Physics, Chinese Academy of
Sciences, P.O. Box 31, Lanzhou 730000, P.R.China\\2. Center of
Theoretical Nuclear Physics, National Laboratory of Heavy Ion
Collisions, Lanzhou 730000, P.R.China\\3. Graduate School, Chinese
Academy of Sciences, Beijing 100049, P.R.China}

\begin{abstract}

We study the near-threshold $\eta '$ production mechanism in
nucleon-nucleon and $\pi N$ collisions under the assumption that
sub-threshold resonance $N^{*}(1535)$ is predominant. In an
effective Lagrangian approach which gives a reasonable description
to the $pN \rightarrow pN \eta$ and $\pi p \rightarrow p \eta$
reactions, it is found that t-channel $\pi$ exchange make the
dominate contribution to the $pN \rightarrow pN \eta '$ process, and
a value of 6.5 for the ratio of $\sigma (pn \rightarrow pn \eta ')$
to $\sigma (pp \rightarrow pp \eta ')$ is predicted. A strong
coupling strength of $N^{*}(1535)$ to $\eta ' N$ ($g_{\eta ' NN^*
}^2 /4\pi = 1.1$) is extracted from a combined analysis to $pp
\rightarrow pp \eta '$ and $\pi N \rightarrow N \eta '$, and the
possible implication to the intrinsic component of $N^{*}(1535)$ is
explored.

\end{abstract}

 \maketitle{}

\section{INTRODUCTION}

As the members of the nonet of the lightest pseudoscalar mesons, the
$\eta$ and $\eta '$ mesons have been the subject of considerable
interest since accurate and complete measurements have been
performed at the experimental facilities of COSY, MAMI, DISTO,
GRAAL, CELSIUS and SATURNE in the past few years. Their intrinsic
structure and properties , as well as the production mechanism in
elementary particle and hadron physics, are intensively explored.
The physically observed $\eta$ and $\eta '$ mesons are mixtures of
the pseudoscalar octet and singlet, which results in a considerable
amount of $s\bar{s}$ in both and accounts for the difference in
$\eta$ mass from the pion. The much greater mass of $\eta '$ meson
is thought to be induced by the non-perturbative gluon
dynamics\cite{gluondy} and the axial anomaly\cite{axial}.

The $\eta$ and $\eta '$ production in nucleon-nucleon collisions
strengthen our understanding on those problems and also provide
assistant opportunities to study the possible nucleon resonances
$N^{*}$ that couple only weakly to pion. Due to the precise
measurements of the total cross section of the $pp \rightarrow pp
\eta$
reaction\cite{ppetasection,Hibou,ppetainva1,ppetainva2,ppetainva4},
a number of
studies\cite{ppetaearly,Wilkin,Riska,Nakappeta1,Nakappeta2,Baru,Ceci,Shyam,Moskal}
have concluded that $\eta$ meson is dominantly produced through the
excitation and de-excitation of the $N^{*}(1535)$ resonance in this
reaction, though the excitation mechanism is still under debate. The
first measurement of the cross section of the quasi-free $pn
\rightarrow pn \eta$ reaction\cite{pnetadata} shows about a factor
of 6.5 larger than that of $pp \rightarrow pp \eta$, clearly
indicating a dominance of isovector exchange. A recent experimental
study of the analyzing power of the $\vec{p} p \rightarrow pp \eta$
reaction\cite{Czyzykiewicz} support that the $\pi$ meson exchange
between the colliding nucleons is predominant. On the other hand,
for the lack of experimentally established baryonic resonances which
would decay into $\eta '$, our understanding of the $\eta '$
production is still much poorer and unsatisfactory, and there are
only a limited number of studies both
experimentally\cite{Hibou,ppetapridata,Balestra,Khoukaz,etapriphodata}
and
theoretically\cite{ppetapriearly,Gedalin,Nakaetapriearly,Nakaetapri}.
An early analysis based on the covariant one Boson exchange(OBE)
model\cite{Gedalin} reproduces the near-threshold total cross
section of the $pp \rightarrow pp \eta '$ reaction without any
resonant term. However, a relativistic meson exchange
model\cite{Nakaetapriearly} demonstrates that the existing data
could be explained either by mesonic and nucleonic currents or by a
dominance of two missing resonances $S_{11}(1897)$ and
$P_{11}(1986)$. The extended study\cite{Nakaetapri} motivated by the
updated data of the $\gamma p \rightarrow \eta '
p$\cite{etapriphodata} and $pp \rightarrow pp \eta
'$\cite{Balestra,Khoukaz} yields resonances $S_{11}(1650)$ and
$P_{11}(1870)$, and it is premature to identify these states, as
these authors pointed out. Besides, another complication comes from
the gluon-induced contact term\cite{Bass}, which would have extra
contribution to the cross-section for $pp \rightarrow pp \eta '$,
since it is possible that $\eta '$ meson couples strongly to gluons.

Recently, high-precision data of the reaction $\gamma p \rightarrow
\eta ' p$ for photon energies from 1.527GeV to 2.227GeV are obtained
by the CLAS Collaboration\cite{Dugger}, and the
analysis\cite{Dugger,Nakaetapriphoto} of these data suggest for the
first time that both the $N^{*}(1535)$ and $N^{*}(1710)$ resonances,
known to couple strongly to the $\eta N$ channel, couple to the
$\eta ' N$ channel. This is obviously the evidence for the important
role of these resonances in the $\eta '$ production. Theoretically,
$N^{*}(1535)$ is found to be important for the near-threshold
$\Lambda$ and $\phi$ production in nucleon-nucleon
collisions\cite{zou}, and a significant coupling of $N^{*}(1535)$ to
strange particles is indicated. Furthermore, the properties of
$N^{*}(1535)$ resonance are extensively discussed in chiral unitary
approach\cite{Kaiser}, and large couplings to $\eta N$, $K \Sigma$
and $K \Lambda$ are also illustrated.

Motivated by these research, in this paper we assume that the
excitation and de-excitation of the $N^{*}(1535)$ resonance play a
major role in the $\eta '$ production in the near-threshold region,
and perform a consistent analysis to the reactions $pp \rightarrow
pp \eta (\eta ')$, $pn \rightarrow pn \eta (\eta ')$ and $\pi N
\rightarrow N \eta (\eta ')$ in the framework of an effective
lagrangian approach. Because the coupling strength of $\eta '$ meson
to the nucleon and $N^{*}$ are poorly
known\cite{Nakaetapri,Dugger,Nakaetapriphoto}, in our analysis we do
not include $N^{*}(1650)$ and $N^{*}(1710)$, which are expected to
have very small contribution to the considered energy
region\cite{Shyam}. The inclusion of the nucleonic and mesonic
currents in the intermediate state is found to make negligible
difference in the final results\cite{Nakappeta2,Shyam}, so we do not
consider them either.

\section{Effective Lagrangian Approach}

We treat the reactions $pp \rightarrow pp \eta (\eta ')$ and $\pi N
\rightarrow N \eta (\eta ')$ at the relativistic tree level in an
effective Lagrangian approach, as depicted by Feynman diagrams in
Fig. 1. Mesons exchanged are restricted to those observed in the
decay channels of the adopted resonances, and most values of the
coupling constants are fixed by the experimental decay ratios. As a
result, the only adjustable parameters are cut-off parameters in the
form factors. All interference terms between different amplitudes
are neglected because the relative phases of these amplitudes are
not known. The relevant meson-nucleon-nucleon(MNN) and
meson-nucleon-resonance(MNR) effective Lagrangians for evaluating
the Feynman diagrams in Fig. 1 are\cite{Tsushima,zou}:
\begin{equation}
L_{\pi NN}  =  - ig_{\pi NN} \bar N\gamma _5 \vec \tau  \cdot \vec
\pi N ,
\end{equation}
\begin{equation}
L_{\rho NN}  =  - g_{\rho NN} \bar N(\gamma _\mu   + \frac{\kappa
}{{2m_N }}\sigma _{\mu \nu } \partial ^\nu  )\vec \tau  \cdot \vec
\rho ^\mu  N ,
\end{equation}
\begin{equation}
L_{\eta NN}  =  - ig_{\eta NN} \bar N\gamma _5 N\eta ,
\end{equation}
\begin{equation}
L_{\pi NN^*}  =  - g_{\pi NN^*} \bar N^* \gamma _5 \vec \tau  \cdot
\vec \pi N^*  + h.c. ,
\end{equation}
\begin{equation}
L_{\rho NN^*}  = ig_{\rho NN^*} \bar N^* \gamma _5 (\gamma _\mu   -
\frac{{q_\mu \gamma \cdot q}}{{q^2 }})\vec \tau  \cdot \vec \rho
^\mu N^* + h.c. ,
\end{equation}
\begin{equation}
L_{\eta NN^*}  =  - g_{\eta NN^*} \bar N^* N^* \eta  + h.c. ,
\end{equation}
\begin{equation}
L_{\eta ' NN^*}  =  - g_{\eta ' NN^*} \bar N^* N^* \eta '  + h.c.
\end{equation}
with $g_{\pi NN}^2 /4\pi = 14.4$, $g_{\rho NN}^2 /4\pi = 0.9$, and
${\kappa  = 6.1}$. The coupling constant $g_{\eta NN}$ is
undetermined nowadays, and the value of $g_{\eta NN}^2 /4\pi$ used
in literature is ranging from 0.25 to 7\cite{Bonn}. Recent
calculations\cite{Wilkin,Ceci,Riska,Baru,zou} seem to favor small
$g_{\eta NN}$, and $g_{\eta NN}^2 /4\pi$ = 0.4\cite{zou} are used in
our calculation. The partial decay width of $N^{*}(1535) \to N\pi$,
$N^{*}(1535) \to N\rho \to N\pi\pi$ and  $N^{*}(1535) \to N\eta$
then can be calculated by above Lagrangians, and the coupling
constants $g_{\pi NN^* }^2 /4\pi$, $g_{\rho NN^* }^2 /4\pi$, and
$g_{\eta NN^* }^2 /4\pi$ are determined through the empirical
branching ratios\cite{Tsushima,zou}, as summarized in Table I. Up to
now, we have no information on the coupling constant of the $\eta '
NN^{*}(1535)$ vertex, and we determine it from a combined analysis
of $pp \rightarrow pp \eta '$ and $\pi N \rightarrow N \eta '$
reactions.

In order to dampen out high values of the exchanged momentum, the
resulting vertexes are multiplied by off-shell form factors. In $pp
\rightarrow pp \eta (\eta ')$ reactions, the form factors used in
the Bonn model\cite{Bonn} are taken:
\begin{equation}
F_M (q^2) = \left(\frac{{\Lambda _M^2  - m_M^2 }}{{\Lambda _M^2  -
q_M^2 }}\right) ^n ,
\end{equation}
with $\Lambda _M$, $q_M$ and $m_M$ being the cut-off parameter,
four-momentum and mass of the exchanged meson. The commonly used
$n=2$ for $\rho NN$ vertex, and $n=1$ for other vertexes, are
employed. The cut-off parameters $\Lambda _{\pi}$ = 1.05GeV for $\pi
NN$, $\Lambda _{\rho}$ = 0.92GeV for $\rho NN$, $\Lambda _{\eta}$ =
2.00GeV for $\eta NN$ and $\Lambda _{M}$ = 0.80GeV for MNR vertexes
are adopted from Ref.\cite{Tsushima}, which performed a systematic
consistent investigation of the strangeness production process in
nucleon-nucleon collisions. In $\pi N \rightarrow N \eta (\eta ')$
reactions, the following form factors for $N^{*}(1535)$ resonance
are used\cite{Nakaetapriearly,Nakaetapri,Nakaetapriphoto,zou}:
\begin{equation}
F_{N^* } (q^2 ) = \frac{{\Lambda ^4 }}{{\Lambda ^4  + (q^2  -
M_{N^*}^2 )^2 }} ,
\end{equation}
with the cut-off parameter $\Lambda$ = 2GeV.

Propagators of $\pi$($\eta$), $\rho$ and $N^{*}(1535)$ are:
\begin{equation}
G_M (q_M ) = \frac{i}{{q_M^2  - m_M^2 }} ,
\end{equation}
\begin{equation}
G_\rho ^{\mu \nu } (q_\rho  ) =  - i\frac{{g^{\mu \nu }  - q_\rho
^\mu  q_\rho ^\nu  /q^2 }}{{q_\rho ^2  - m_\rho ^2 }} ,
\end{equation}
\begin{equation}
G_R (p_R ) = \frac{{\gamma \cdot p_R + m_R }}{{p_R^2  - m_R^2 + im_R
\Gamma _R }} .
\end{equation}

With above formalism, the invariant amplitude can be obtained
straightforwardly by applying the Feynman rules to Fig. 1.

It is generally agreed that $^{1}S_{0}$ proton-proton final state
interaction (FSI) influences the near-threshold behavior
significantly in $pp \rightarrow pp \eta(\eta ')$. In present
calculation, Watson-Migdal factorization\cite{FSI} are used and the
pp FSI enhancement factor is taken to be Jost function\cite{Jost}:
\begin{equation}
\left| {J(k)} \right|^{ - 1}  = \frac{{k + i\beta }}{{k - i\alpha }}
.
\end{equation}
where $k$ is the internal momentum of $pp$ subsystem. The related
scattering length and effective range are:
\begin{equation}
a = \frac{{\alpha  + \beta }}{{\alpha \beta }} ,
\begin{array}{*{20}c}
   {} & {}  \\
\end{array}
r =\frac{2}{{\alpha  + \beta }} \label{eq:fsia} ,
\end{equation}
with $a$ = -7.82fm and $r$ = 2.79fm(i.e. $\alpha$ = -20.5MeV and
$\beta$ = 166.7MeV) for $^{1}S_{0}$ $pp$ interaction.

Then the total cross section can be calculated by above
prescription, and the integration over the phase space can be
performed by Monte Carlo program. As to the $pn \rightarrow pn \eta
(\eta ')$ reaction, isospin factors are
considered\cite{Wilkin,Nakappeta1,Shyam}, and $a$ = -23.76fm and $r$
= 2.75fm(i.e. $\alpha$ = -7.87MeV and $\beta$ = 151.4MeV) for
$^{1}S_{0}$ $pn$ interaction, $a$ = 5.424fm and $r$ = 1.759fm(i.e.
$\alpha$ = 45.7MeV and $\beta$ = 178.7MeV) for $^{3}S_{1}$ $pn$
interaction are used.

\section{Numerical Results}

We first apply our approach to the $\eta$ production, and check the
applicability of our model. Total cross section for $pp \rightarrow
pp \eta$, $\pi ^{-} p \rightarrow n \eta $ and $pn \rightarrow pn
\eta$ are shown in Fig. 2, and our numerical results agree well with
the experimental data. Contributions of various meson exchanges to
$pp \rightarrow pp \eta$ and $pn \rightarrow pn \eta$ are also
shown, and $\pi$ exchange is found to make dominant contribution in
the near-threshold region. This has received support from recent
experiment\cite{Czyzykiewicz}, and also the reason for our
simultaneous reproduce to these two
channels\cite{Nakappeta1,Shyam,pnetadata}. In sharp contrast to
Ref.\cite{Wilkin} which indicates $\rho$ exchange dominance, the
contribution of $\rho$ exchange is much smaller than that of $\pi$
and $\eta$ exchange in our calculation. Besides, in a
calculation\cite{zou} to $pp \rightarrow pp \phi$ reaction whose
approach is similar to us, it is demonstrated that the contribution
of $\rho$ exchange is larger than that of $\eta$ exchange though
$\pi$ exchange is dominant in the $N^{*}(1535)$ excitation. This
difference to our model is caused by the alternative cut-off
parameters in the form factors, and much larger
values($\Lambda=1.6GeV$ for $\rho NN$ vertex and $\Lambda=1.3GeV$
for all other form factors) are used in their model. It seems that
the vector couplings of the $\rho NN$ vertex are suppressed more
fast than the pseudo-scalar couplings of $\pi NN$ and $\eta NN$
vertex when the cut-off parameters are decreased. In the considered
energy region, the small cut-off parameters should be more
reasonable, as already illustrated in the analysis to the
strangeness production process in nucleon-nucleon
collisions\cite{Tsushima}. Similarly, our model should draw some
analogous conclusions to the $pN \rightarrow pN \eta '$ channel in
this aspect due to the formalism of our model, as demonstrated
below. The relatively larger $\eta$ exchange contribution than that
of $\rho$ exchange is also found in Refs.\cite{Nakappeta1,Ceci}, but
it is worth pointing out that a very small $g_{\eta NN}$ is adopted
in our model.

As can be seen from Fig. 2(a)(c), there is no much room left to the
coherent resonance-resonance interference term, which is thought to
be non-negligible, as stressed in Ref.\cite{Shyam}. The cross
section of $\pi ^{-} p \rightarrow n \eta$ where $T_{\pi}
>$ 850MeV is underestimated as displayed in Fig. 2(b), and this is
obviously the evidence to the contribution of other resonances(i.e.
$N^{*}(1650)$ and $N^{*}(1710)$) in this energy region.

For excess energies smaller than 20MeV, theoretical results
underestimate the empirical cross section of $pp \rightarrow pp
\eta$ channel, as several authors pointed
out\cite{Nakappeta1,Shyam}. The discrepancy in invariant mass
distribution is even more pronounced, as can be clearly seen in Fig.
3(a-d). In addition to a peak arising from the $N^{*}(1535)$
resonance and strong $^{1}S_{0}$ pp FSI, there is a surprising broad
bump in both $pp$ and $p \eta$ invariant mass distribution, which is
not trivial to be explained. Some papers devote to this problem, and
the origin of the bump is attributed to the large $\eta$ meson
exchange contribution comparable with the leading $\pi$ meson
exchange term\cite{Ceci} or higher partial waves\cite{Nakappeta2}.
However, former hypothesis apparently conflicts with the
experimental finding of a dominance of isovector exchange, thus it
can not account for the high ratio of $\sigma (pn \rightarrow pn
\eta)$ to $\sigma (pp \rightarrow pp \eta)$. The latter can not give
simultaneous explanation of the excitation function and invariant
mass distributions, and the visible bump at excess energies of
4.5MeV\cite{ppetainva4} is either improbably caused by the
contribution of higher partial waves. As a result, it seems that
this bump probably arises from the $\eta N$ FSI\cite{Moskal}.
Unfortunately, till now there is no rigorous treatment of three-pair
FSI, and this problem needs further theoretical and experimental
effort. As shown in Fig. 3(e-f), the angular distribution of $\eta$
meson in the $pp \rightarrow pp \eta$ reaction for excess energies
of 15MeV and 41MeV are described well by our model, since our model
is characterized by the $\pi$ exchange dominance process in the
$N^{*}(1535)$ excitation.

Then we will employ our model to $\eta '$ production since its
success to $\eta$ production has been demonstrated above. Total
cross section for $pp \rightarrow pp \eta '$, $\pi N \rightarrow N
\eta '$ and $pn \rightarrow pn \eta '$ are shown in Fig. 4. We get
good reproduce to both $pp \rightarrow pp \eta '$ and $\pi N
\rightarrow N \eta '$ channels with $g_{\eta ' NN^* }^2 /4\pi$ =
1.1, and some similar conclusions to $\eta$ production are achieved
as expected. $\pi$ exchange is the largest contribution in the
near-threshold region of $pN \rightarrow pN \eta '$, and $\rho$
exchange is much smaller than $\pi$ and $\eta$ exchange. Without
complexity caused by $\eta ' N$
interaction\cite{ppetapridata,ppetapriearly}, our numerical results
reproduce the experimental data quite well in the whole considered
energy region. As can be seen in Fig. 4(c), we anticipate the same
value of 6.5 for the ratio of $\sigma (pn \rightarrow pn \eta ')$ to
$\sigma (pp \rightarrow pp \eta ')$ in our model, while this ratio
will approach unity if $\eta '$ is produced directly by
gluons\cite{Bass}. So isospin dependence is powerful to distinguish
different $\eta '$ production mechanism, and may provide useful
information to the possible gluon content of $\eta '$ meson.

For the scarce and inaccurate data of $\pi N \rightarrow N \eta '$,
the extracted coupling constant $g_{\eta ' NN^* }$ has large err
bar, and significant contributions from other $N^{*}$ resonances
cannot be definitely excluded. Alternative combination of $N^{*}$
resonances and coupling strength would yield a good fit to present
data\cite{zou}. The dotted curve in Fig. 4(b) shows that we can get
a much better reproduce to the $\pi N \rightarrow N \eta '$ data
with $g_{\eta ' NN^* }^2 /4\pi$ = 1.0, although this will slightly
underestimate the $pp \rightarrow pp \eta '$ channel. An even better
fit to the $pp \rightarrow pp \eta '$ data can be achieved with
$g_{\eta ' NN^* }^2 /4\pi$ = 1.15, but this will overestimate the
$\pi N \rightarrow N \eta '$ data as shown by the dashed line in Fig
4(b). Anyway, we get good result to both channels with $g_{\eta '
NN^* }^2 /4\pi$ = 1.1, and our preliminary analysis should be
reasonable considering that other $N^{*}$ resonances except
$N^{*}(1535)$ show very weak couplings to $\eta ' N$.

The calculated invariant mass spectrum of $pp \rightarrow pp \eta '$
reaction at the excess energies of 15.5MeV, 46.6MeV and 143.8MeV are
presented in Fig. 5. Our calculations of angular distribution of
$\eta '$ meson at 46.6MeV and 143.8MeV show obvious structure at
forward and backward angles, and reproduce the experimental data
nicely. However, it has to admitted that the measured angular
dependence might be also compatible to isotropic shape within the
given experimental uncertainties. Besides, it is interesting to note
that the data from Ref.\cite{ppetainva1} show distinct structure in
the angular distribution of $\eta$ meson, but Ref.\cite{ppetainva2}
gives a totally flat distribution, as can be seen in Fig. 3(e-f). So
a detailed quantitative analysis awaits for the clearing of the
experimental situation.

The predicted differential cross section of $pp \rightarrow pp \eta
'$ at the excess energy of 15.5MeV, together with the total cross
section of $pn \rightarrow pn \eta '$, can be examined by the
ongoing experimental studies\cite{Moskal}. No obvious bump other
than a peak arise in the invariant mass distribution because our
model do not include additional mechanism rather than the
$N^{*}(1535)$ resonance and FSI. If this is confirmed by the
experiment, then other mechanism(probably the $\eta N$ FSI)
accounting for the broad bump should be added to the study of the
$pp \rightarrow pp \eta$ channel.

\section{Summary and Discussion}

In this paper, we present a consistent analysis to $pN \rightarrow
pN \eta '$ and $\pi N \rightarrow N \eta '$ within an effective
Lagrangian approach, assuming that $N^{*}(1535)$ resonance is
dominant in the $\eta '$ production. Our numerical results show that
$\pi$ exchange is the most important in $pN \rightarrow pN \eta '$
reaction, and predict a large ratio of $\sigma (pn \rightarrow pn
\eta ')$ to $\sigma (pp \rightarrow pp \eta ')$. An explicit
structure in angular distribution of $\eta '$ meson is demonstrated.
Besides, a significant coupling strength of $N^{*}(1535)$ to $\eta '
N$ is found:
\begin{equation}
g_{\eta ' NN^* }^2 /4\pi = 1.1
\end{equation}
In a vector-meson-dominant model analysis to $\gamma p \rightarrow p
\eta '$ reaction\cite{Sibirtsev}, a value of $g_{\eta ' NN^* } =
3.4$(i.e. $g_{\eta ' NN^* }^2 /4\pi = 0.92$) is given, and this is
coincident to our analysis. We would illustrate that this is also
compatible to the mixture picture of $\eta$ and $\eta '$.

Considering the possible gluonium admixture of the $\eta '$ wave
function, a basis of states $|\eta _{q} \rangle = |u\bar{u} +
d\bar{d} \rangle/\sqrt{2}$, $|\eta _{s} \rangle = |s\bar{s} \rangle$
and $|G \rangle = |Gluonium \rangle$ is adopted, and the physical
$\eta $ and $\eta '$ are assumed to be linear combinations of these
basis of states\cite{gluon,nogluon}:
\begin{equation}
|\eta \rangle = X_{\eta} |\eta _{q} \rangle +  Y_{\eta} |\eta _{s}
\rangle + Z_{\eta} |G \rangle
\end{equation}
\begin{equation}
|\eta ' \rangle = X_{\eta '} |\eta _{q} \rangle +  Y_{\eta '} |\eta
_{s} \rangle + Z_{\eta '} |G \rangle
\end{equation}
If the gluonium content of the $\eta$ meson is assumed to
vanish($Z_{\eta} = 0$), all six parameters can be written in terms
of two mixing angles, $\phi _{p}$ and $\phi _{\eta 'G}$, which
correspond to:
\begin{equation}
X_{\eta} = cos\phi _{p} ,
\begin{array}{*{20}c}
   {} & {}  \\
\end{array}
Y_{\eta} = -sin\phi _{p},
\begin{array}{*{20}c}
   {} & {}  \\
\end{array}
Z_{\eta} = 0 ,
\end{equation}
\begin{equation}
X_{\eta '} = sin\phi _{p} cos\phi _{\eta 'G} ,
\begin{array}{*{20}c}
   {} & {}  \\
\end{array}
Y_{\eta '} = cos\phi _{p} cos\phi _{\eta 'G} ,
\begin{array}{*{20}c}
   {} & {}  \\
\end{array}
Z_{\eta '} = -sin\phi _{\eta 'G} .
\end{equation}
If  the gluonium content of the $\eta '$ meson is further assumed to
vanish($Z_{\eta} = 0$, i.e. $\phi _{\eta 'G} = 0$), then $\phi _{p}$
is the $\eta - \eta '$ mixing angle in absence of gluonium, and
Eqs.(16)(17) are the normal $\eta - \eta '$ mixing in the
quark-flavor basis. In the quark model, the $\eta '$ couplings can
be related to those of $\eta$\cite{Gedalin,Sibirtsev}:
\begin{equation}
g_{\eta} = X_{\eta} g_{q} +  Y_{\eta} g_{s} + Z_{\eta} g_{G}
\end{equation}
\begin{equation}
g_{\eta '} = X_{\eta '} g_{q} +  Y_{\eta '} g_{s} + Z_{\eta '} g_{G}
\end{equation}
with $g_{q}$, $g_{s}$ and $g_{G}$ being the non-strangeness,
strangeness and gluonium coupling constant. As to $g_{\eta ' NN}$
and $g_{\eta NN}$, because the strangeness and gluonium content in
nucleon are negligible, we can take the simplifying assumption
$g_{s} \ll g_{q}$ and $g_{G} \ll g_{q}$:
\begin{equation}
R_{N}=\frac{g_{\eta ' NN}}{g_{\eta NN}} \simeq \frac{X_{\eta
'}}{X_{\eta}} = tan\phi _{p} \sim 0.84
\end{equation}
with $\phi _{p} \sim 40^{\circ}$\cite{mixingangle}. This is
compatible to $R_{N} \sim 0.62$ with recently extracted value of
$g_{\eta ' NN} \simeq 1.4$\cite{Dugger} and adopted $g_{\eta NN}^2
/4\pi = 0.4$ in this paper.

With coupling constants summarized in Table. I, we have:
\begin{equation}
R_{N^*}=\frac{g_{\eta ' NN^*}}{g_{\eta NN^*}} \sim 2.0
\end{equation}
If the large $g_{\eta ' NN^*}$ indeed indicates a significant
$s\bar{s}$ configuration inside $N^{*}(1535)$ resonance\cite{zou},
assuming $g_{GNN^*} \ll g_{qNN^*}$ should be reasonable:
\begin{equation}
R_{N^*}=\frac{g_{\eta ' NN^*}}{g_{\eta NN^*}} = \frac{tan\phi _{p} +
g_{sNN^*}/g_{qNN^*}}{1 - g_{sNN^*}/g_{qNN^*}tan\phi _{p}}
\end{equation}
Then we will get $g_{sNN^*}/g_{qNN^*} \sim 0.43$, which may indicate
a relatively large proportion of strangeness in $N^{*}(1535)$
resonance. But the large $g_{\eta ' NN^*}$ is also probably caused
by the gluonium component of $N^{*}(1535)$ as can be seen in
Eqs.(20)(21), then if $g_{sNN^*} \ll g_{qNN^*}$ is assumed:
\begin{equation}
R_{N^*}=\frac{g_{\eta ' NN^*}}{g_{\eta NN^*}} = tan\phi _{p}cos\phi
_{\eta 'G} - \frac{g_{GNN^*}}{g_{qNN^*}} \frac{sin\phi _{\eta
'G}}{cos\phi _{p}}
\end{equation}
where $\phi _{p} \sim 40^\circ$ and $|\phi _{\eta 'G}| \sim
22^\circ$\cite{gluon}. Then we will get $|g_{GNN^*}/g_{qNN^*}| \sim
2.5$, and this may also indicate a relatively large proportion of
gluons in $N^{*}(1535)$ resonance. Certainly, according to above
analysis, it is possible that strangeness and gluons coexist in
$N^{*}(1535)$, and it is two of them that induce the large couplings
of $N^{*}(1535)$ to strange particles. Recently, phenomenological
analysis of radiative decays and other processes\cite{nogluon}
conclude no evidence of the gluonium contribution of $\eta '$ wave
function(i.e. $|\phi _{\eta 'G}| \sim 0^\circ$), and this seems to
support the idea that these large couplings are major caused by the
$s\bar{s}$ component in $N^{*}(1535)$. Different 5-quark
configurations of $qqqs\bar{s}$ have deeply investigated, and
admixture of 25-65\% in $N^{*}(1535)$ is suggested\cite{CSAn}.
However, the intrinsic structure of $N^{*}(1535)$ is still left to
be an open question and further study are needed.

In conclusion, our phenomenological analysis to the $\eta '$
production in nucleon-nucleon and $\pi N$ collisions not only give
nice reproduce to the experimental data, but also agree well with
the present understanding of the internal component of the $\eta$
($\eta '$) meson and $N^{*}(1535)$ resonance, although alternative
contribution from other $N^{*}$ resonances are also possible. The
ongoing relevant experiment in COSY\cite{Moskal} will soon examine
our results and advance a better knowledge of the $\eta$ and $\eta
'$ production.

\begin{acknowledgments}

We would like to thank Q. W. Wang, J. J. Xie and B. S. Zou for
fruitful discussions and program code. This work was supported by
the CAS Knowledge Innovation Project (No.KJCX3-SYW-N2,
No.KJCX2-SW-N16) and Science Foundation of China (10435080,
10575123, 10710172).

\end{acknowledgments}

\begin{table}
\caption{Relevant $N^* (1535)$ parameters} \label{table1}
\begin{center}
\begin{tabular}{c  c c c c c c c}
\hline\hline
 &     & Width    & Channel    & Branching ratio    & Adopted value    & $g^2 /4\pi $ &
\\
\hline & $N^* (1535)$ &150MeV & $\pi N$ & 0.35-0.55 & 0.45 & 0.033 &
\\
&   &   & $\rho N$ & 0.02$\pm$ 0.01&  0.02 & 0.10  &\\
&   &   & $\eta N$ & 0.45-0.60&  0.53 & 0.28  &\\
&   &   & $\eta ' N$ & ---&  --- & 1.1  &\\
\hline \label{table1}
\end{tabular}
\end{center}
\end{table}

\begin{figure}[ht]
  \begin{center}
    \rotatebox{0}{\includegraphics*[width=0.7\textwidth]{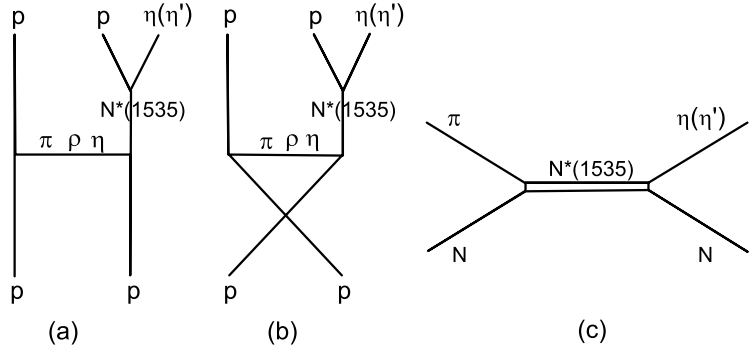}}
    \caption{Feynman diagrams for $pp \rightarrow pp \eta (\eta ')$ and $\pi N \rightarrow N \eta (\eta ')$.}
  \end{center}
\end{figure}

\begin{figure}[ht]
  \begin{center}
    \rotatebox{0}{\includegraphics*[width=0.7\textwidth]{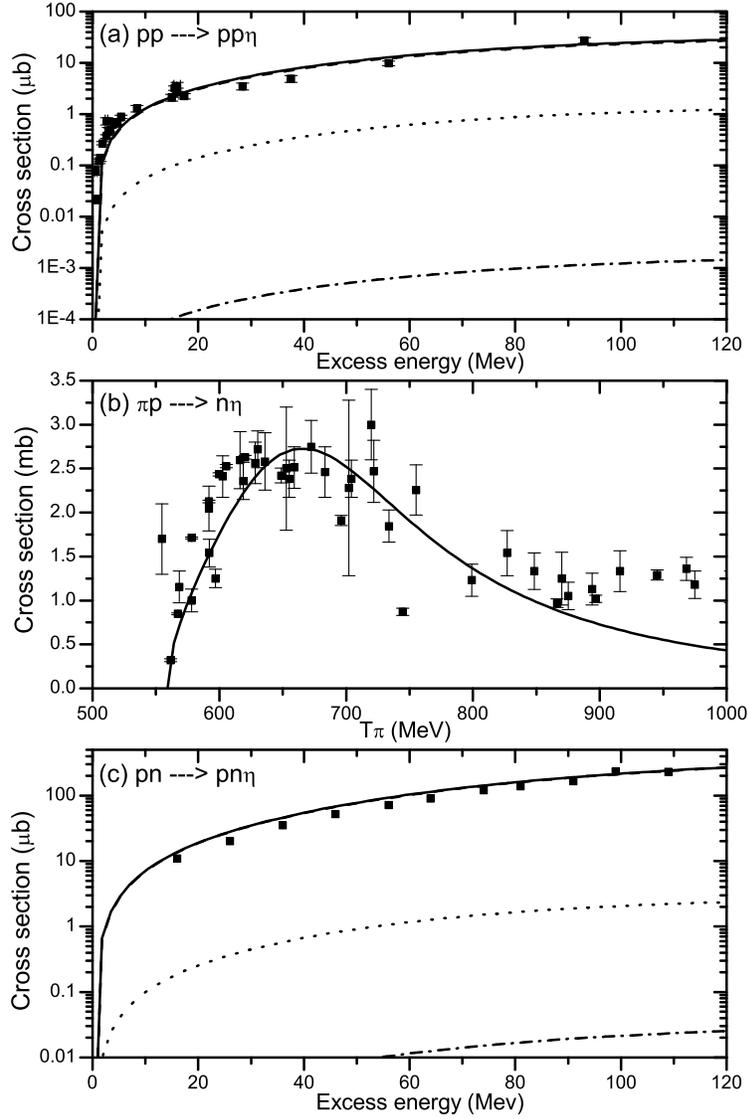}}
    \caption{Total cross section for $pp \rightarrow pp \eta$(a), $\pi ^{-} p \rightarrow n \eta$(b) and $pn \rightarrow pn \eta$(c). (a)(c): The dashed, dotted, dash-dotted and solid curve correspond to contribution from $\pi$, $\eta$, $\rho$ exchange and their simple sum, respectively. The dashed curve is overlapped by the solid one. The data are from Ref.\protect\cite{ppetasection,Hibou}(a), Ref.\protect\cite{databook}(b) and
    Ref.\protect\cite{pnetadata}(c).}
  \end{center}
\end{figure}

\begin{figure}[ht]
  \begin{center}
    \rotatebox{0}{\includegraphics*[width=0.7\textwidth]{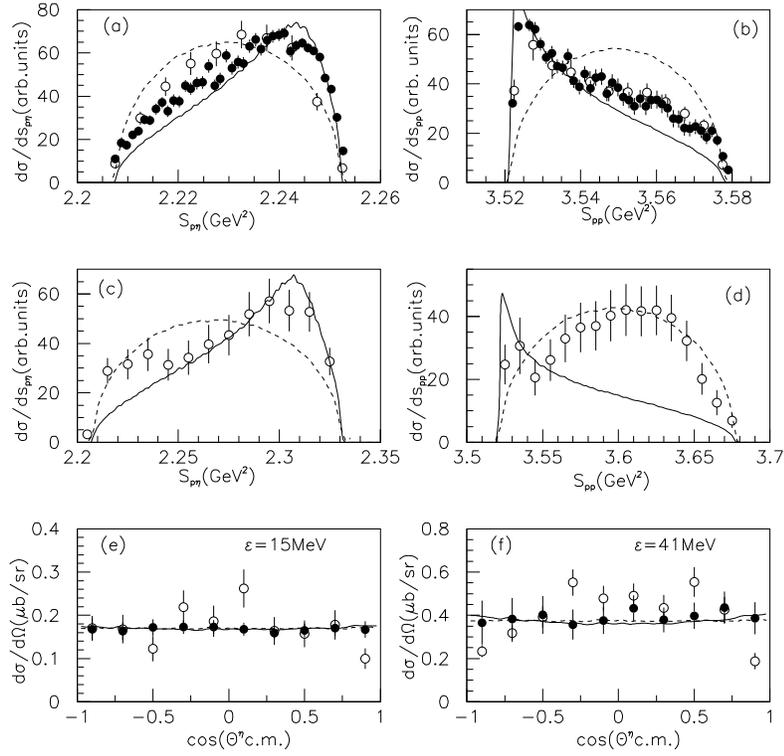}}
  \caption{Invariant mass spectrum for $pp \rightarrow pp \eta$. (a)(b)(e) and (c)(d)(f) are invariant mass spectrum at excess energies of 15MeV and 41MeV, respectively. (a-d): The data are from Ref.\protect\cite{ppetainva2}(open circle) and Ref.\protect\cite{ppetainva4}(closed circle). (e-f): The data are from Ref.\protect\cite{ppetainva1}(open circle) and Ref.\protect\cite{ppetainva2}(closed circle). The dashed curve is the pure phase-space distribution.}
  \end{center}
\end{figure}

\begin{figure}[ht]
  \begin{center}
    \rotatebox{0}{\includegraphics*[width=0.7\textwidth]{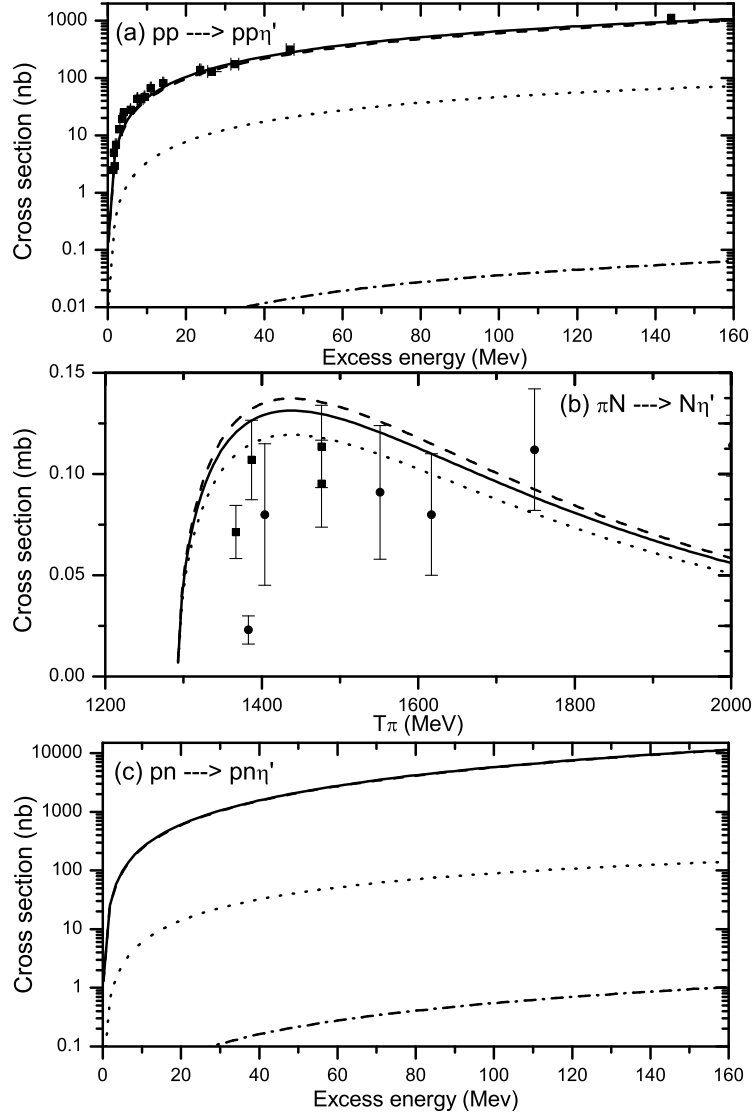}}
  \caption{Total cross section for $pp \rightarrow pp \eta '$(a), $\pi N \rightarrow N \eta '$(b) and $pn \rightarrow pn \eta '$(c). (a)(c): Same as Fig. 2(a)(c). (b): The dashed, solid and dotted curve correspond to $g_{\eta ' NN^* }^2 /4\pi$ = 1.15, 1.1 and 1.0. The data are from Ref.\protect\cite{ppetapridata}(a),
  Ref.\protect\cite{databook}(b)(closed square: $\pi ^{-} p \rightarrow n \eta '$, closed circle: $\pi ^{+} n \rightarrow p \eta '$).}
  \end{center}
\end{figure}

\begin{figure}[ht]
  \begin{center}
    \rotatebox{0}{\includegraphics*[width=0.7\textwidth]{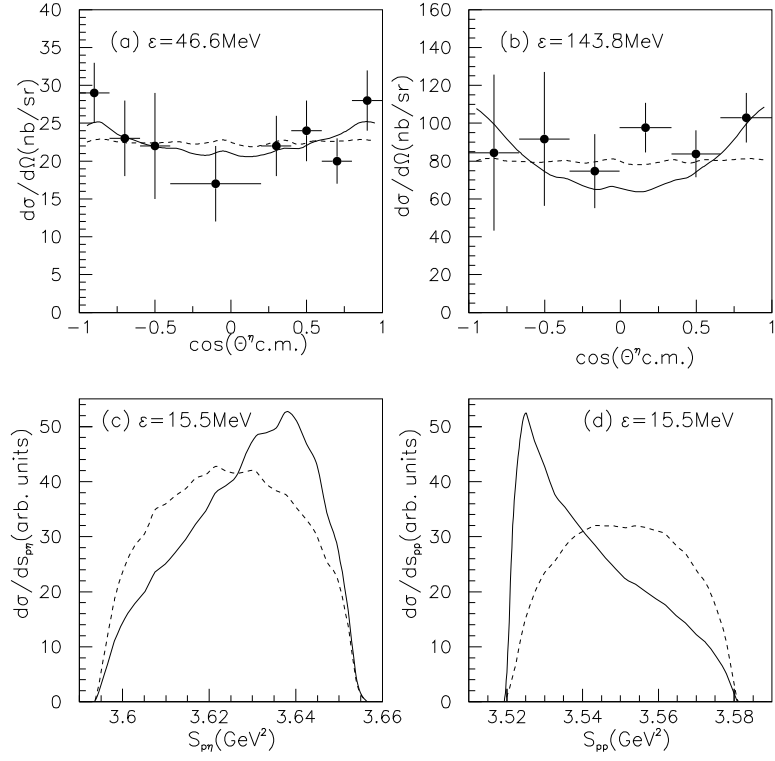}}
  \caption{Invariant mass spectrum for $pp \rightarrow pp \eta '$. (a)(b) and (c)(d) are angular distribution of $\eta$ meson and invariant mass distribution respectively. The data are from Ref.\protect\cite{Khoukaz}(a) and Ref.\protect\cite{Balestra}(b). The dashed curve is the pure phase-space
  distribution.}
  \end{center}
\end{figure}

\end{document}